\documentstyle{article}
\def\emline#1#2#3#4#5#6{%
       \put(#1,#2){\special{em:moveto}}%
       \put(#4,#5){\special{em:lineto}}}
\def\newpic#1{}
\date{}
\newcommand{\be}{\begin{equation}}
\newcommand{\ee}{\end{equation}}
\newcommand{\ba}{\begin{array}}
\newcommand{\ea}{\end{array}}
\newtheorem{dfn}{Definition}
\newtheorem{lemma}{Lemma}
\newtheorem{remark}{Remark}
\title{Vacuum Curves, Classical Integrable Systems in Discrete Space-Time and
Statistical Physics}
\author{I~G~Korepanov\\ {\footnotesize{\it Chelyabinsk, pr Lenina 78-A kv 45,
454080, Russia}}}
\begin{document}
\maketitle
\begin{abstract}
A dynamical system with discrete time is studied by means of algebraic
geometry. The system admits a reduction that is interpreted as a classical
field theory in 2+1-dimensional wholly discrete space-time. The integrals of
motion of a particular case of the reduced system are shown to coincide, in
essence, with the statistical sum of the well-known (inhomogeneous)
2-dimensional dimer model (the statistical sum is here a function of two
parameters). Possible generalizations of the system are examined.
\end{abstract}

Vacuum curves and vacuum vectors are algebro-geometrical objects that have
arisen in the theory of the quantum Yang---Baxter equation. They seem to have
their origin in R.~Baxter's works \cite{Baxter1,Baxter2}, and their
general definition was formulated by I.~M.~Krichever \cite{Krichever}.
Baxter, and also Takhtajan and Faddeev \cite{FT} used the vacuum vectors to
obtain a generalization of the Bethe ansatz for the XYZ spin model---%
an integrable one-dimensional quantum field theory (and also for the eight-%
vertex model of two-dimensional statistical physics). Krichever has applied
the vacuum curves to classification of the solutions of the quantum Yang---%
Baxter equation in the tensor product of 2-dimensional vector spaces. Then,
the author of this paper has found some further applications of the vacuum
curves and vacuum vectors. In the paper \cite{Korepanov-vac} (see also
\cite{Korepanov-dis}), new solutions of the quantum Yang---Baxter equation
were constructed for the first time. They correspond to what is now known as
Chiral Potts model. In \cite{Korepanov-hid,Korepanov-spe,%
Korepanov-dis}, the degeneracies of the spectrum of the XXZ
quantum chain hamiltonian were examined by means of the vacuum curves, and in
\cite{Korepanov-tet,Korepanov-dis} the studying of vacuum vector
bundles has resulted in the construction of solutions to the tetrahedron
equation with commuting spin variables on the links.

Here, I try to demonstrate that the vacuum curves may be useful for studying
the classical (not quantum) field theory models as well. A difference
equation on the 2+1-dimensional cubic lattice is presented, for which the
solution to the Cauchy problem is constructed, at least in principle, through
a rather simple scheme. The evolution is of hyperbolic nature, i.e.\ the
``perturbations'' propagate not faster than fixed speed. The interesting
feature is that, in a particular ``scalar'' case, the model reveals a quite
natural connection with the well-known dimer model of statistical physics.
The statistical sum of this latter model, which depends here on two
parameters (and the model itself is, of course, inhomogeneous), is the
integral of motion for any values of these parameters.

This field theory comes as a ``reduction'' of some very simply described
``non-local'' dynamical system. On the other
hand, generalizations of this latter system are constructed in this paper, and
I use a discrete analog of Lax pair for this purpose.

{\bf Acknoledgements.} I owe to A.~B.~Shabat the idea of ``local reduction''
(see Section~\ref{reduction}). L.~D.~Faddeev informed me of the
paper~\cite{BMV}. I would like to express my gratitude to them.

\section{Definition of the dynamical system. Gauge invariance}
\label{gensys}

Let
$$ L=\left(\ba{cc} A&B\\C&D\ea\right)$$
be a block matrix, $A, \ldots D$ being $n\times n$ matrices consisting of
complex numbers. Consider the following two operations: construction of the
inverse matrix
$$L\rightarrow L^{-1}$$
and the block transposing
$$L=\left(\ba{cc} A&B\\C&D\ea\right)\rightarrow L^t=\left(\ba{cc}
A&C\\B&D\ea\right).$$
Now let a (birational) mapping $f$ be a composition of these two operations:
\be f(L)=(L^{-1})^t.\label{1.1}\ee
Let us introduce the discrete integer-valued time $\tau$, and let the matrix
$L$ depend on $\tau$ so that
\be L(\tau+1)=f(L(\tau)).\label{1.2}\ee

This ``dynamical system'' has been already mentioned in literature \cite{BMV}.
In the present paper, the integrability of this system is demonstrated,
assuming that the
``motion'' is considered up to a ``gauge transformation'' (see below).

Let $G$ and $H$ be non-degenerate $n\times n$ matrices. The gauge
transformation of the matrix $L$ is the following transformation of its
blocks:
\be A\rightarrow GAH,\ldots \;D\rightarrow GDH.\label{1.3}\ee
Two matrices $L$ and $L'$ connected by the transformation~(\ref{1.3}) will be
called gauge equivalent. It is clear that if $L(\tau)$ and $L'(\tau)$ belong
to the same class of gauge equivalence, the matrices $L(\tau+1)$ and
$L'(\tau+1)$ also do so. Thus, dynamics~(\ref{1.2}) induces a dynamics on the
set of classes of gauge invariance.

\section{Vacuum curves and vacuum vectors}
\label{vacuum}

It turns out that the dinamics~(\ref{1.2}) preserves the so-called vacuum
curve $\Gamma$ of the operator $L$ (the bases being fixed, we make no
difference between a linear operator and its matrix). To be exact, $\Gamma$
remains unchanged under the transformation $f\circ f$, and undergoes a simple
transformation under $f$. The curve $\Gamma$ together with the class of linear
equivalence of the pole divisor of the vacuum vectors (see below) determines
the matrix $L$ up to a gauge transformation. The set of those classes of
linear equivalence is isomorphic to a complex torus---the Jacobian of the
curve $\Gamma$. The dynamics~(\ref{1.2}) linearizes on the Jacobian, i.e.\
the transformation $f$ corresponds to a constant shift on the torus. Now,
let us discuss these facts in detail.

The vacuum curve of the operator $L$ is an algebraic curve in the space
${\rm C}^2$ of two variables $u,v$. Here are two equivalent definitions of it
\cite{Krichever}.

\begin{dfn}
\label{dfn1}
Consider the relation
\be L(U\otimes X)=V\otimes Y,\label{2.1}\ee
wherein
$$U=\left(\ba{c}u\\1\ea\right),\quad V=\left(\ba{c}v\\1\ea\right)$$
are two-dimensional vectors, $X$ and $\;Y$ are $n$-dimensional vectors. For a
generic matrix $L$, the non-zero solutions $(U,V,X,Y)$ of the
relation~(\ref{2.1}) are parametrized, up to a scalar factor in $X$ and $Y$,
by points of an algebraic curve $\Gamma$ of genus $g=(n-1)^2$ given by an
equation of the form
\be P(u,v)=0,\label{2.2}\ee
$P(u,v)$ being a polynomial of degree $n$ in each variable, i.e.
\be P(u,v)=\sum_{j,k=1}^n a_{jk}u^jv^k.\label{2.3}\ee
$\Gamma$ is called the vacuum curve of the operator $L$.
\end{dfn}
\begin{dfn}
\label{dfn2}
The vacuum curve of the operator $L$ is the curve $\Gamma$ in ${\rm C}^2$
given by the equation
\be P(u,v)=\det(V^\perp LU)=\det(uA+B-uvC-vD)=0,\label{2.4}\ee
where
$$V^\perp=(1,-v).$$
\end{dfn}

Let us denote the points of the vacuum curve by the letter $z=(u,v)\in\Gamma$.
Then $U=U(z)$ and $V=V(z)$ are meromorphic vectors on $\Gamma$ with the pole
divisors $D_U$ and $D_V$ of degree $n$, while $X=X(z)$ and $Y=Y(z)$, if
normalized by, e.g.,\ the condition that their $n$th coordinates equal unity,
become meromorphic vectors with pole divisors $D_X$ and $D_Y$ of degree
$n^2-n$ \cite{Krichever}. Under this normalization, a meromorphic scalar
factor $h(z)$ must be added into~(\ref{2.1}):
\be L(U(z)\otimes X(z))=h(z)V(z)\otimes Y(z).\label{2.5}\ee
The linear equivalence of divisors
$$D_U+D_X\sim D_V+D_Y$$
holds and is provided by the function $h(z)$ in the sense that $h(z)$ has its
poles in the points of $D_U+D_X$ and zeros in the points of $D_V+D_Y$.

As is shown in the paper \cite{Krichever}, the vacuum curve equation $P(u,v)=
0$ and the class of linear equivalence of divisor $D_X$ or $D_Y$ determine a
generic matrix $L$ to within a gauge transformation, and vice versa, the
gauge transformations do not change the vacuum curve and the classes of linear
equivalence of divisors. In other words, the correspondence
$$(\mbox{class of gauge equivalence of }L)\leftrightarrow(\Gamma,\mbox{ the
class of }D_X)$$
is a birational isomorphism.

We will call $X(z)$ the vacuum vector and $Y(z)$ the covacuum vector in the
point $z$ of the curve $\Gamma$. $X(z)=X(u,v)$ generates the (one-dimensional)
kernel of the matrix
\be uA+B-uvC-vD.\label{2.6}\ee

The Definition~\ref{dfn1} allows one to trace what happens with the vacuum
curve and vacuum vactors under the transformation $L\rightarrow L^{-1}$, while
the Definition~\ref{dfn2} allows one to trace what happens under the
transformation $L\rightarrow L^t$. Namely, it is seen from the relation
$$L^{-1}\bigl(V(z)\otimes Y(z)\bigr)=h(z)^{-1}U(z)\otimes X(z)$$
that the vacuum curve equation for the matrix $L^{-1}$ is
$$P(v,u)=0,$$
while its vacuum vector in the point $(v,u)$ coincides with the covacuum
vector of the initial matrix $L$. As for the block transposing, the vacuum
curve equation for the matrix $L^t$
$$\det(uA+C-uvB-vD)=0$$
may be rewritten as
$$u^nv^n\det(v^{-1}A-B+u^{-1}v^{-1}C-u^{-1}D)=0,$$
i.e.
$$u^nv^nP(-v^{-1},-u^{-1})=0.$$
The vacuum vector of the matrix $L^t$ in the point $(-v^{-1},-u^{-1})$ of its
vacuum curve coincides with the vacuum vector $X(u,v)$ of the matrix $L$.

Combining these considerations, one finds out that the vacuum curve $\tilde
\Gamma$ of the matrix $(L^{-1})^t$ is given by equation
$$u^nv^nP(-u^{-1},-v^{-1})=0,$$
while the vacuum vector $\tilde X(-u^{-1},-v^{-1})$ coincides with the vector
$Y(u,v)$ of the matrix $L$.

Identifying the curves $\Gamma$ and $\tilde\Gamma$ by means of the isomorphism
$$(u,v)\leftrightarrow(-u^{-1},-v^{-1}),$$
one sees that
$$D_{\tilde X}\sim D_Y\sim D_X+D_U-D_V,$$
which means that, in essence, the transformation~(\ref{1.1}) results in adding
a fixed element of the Picard group, namely the equivalence class of the
divisor $D_U-D_V$, to the pole divisor $D_X$ of the vacuum vectors. It is
clear also that after two transformations one returns to the initial curve:
$$\tilde{\tilde\Gamma}=\Gamma.$$

\section{Reduction to evolution equation in the 2+1-dimensional space-time}
\label{reduction}

The dynamical system of the previous section admits an interesting reduction,
i.e.\ some special choice of the matrices $A,\ldots D$ that is in agreement
with the evolution. In this section, it will be convenient to treat the
matrices $A,\ldots D$ as linear operators acting from the linear space
${\cal H}_1$ into the linear space ${\cal H}_2$ (of the same finite
dimension). This being the situation at the moment $\tau$, the operators act,
of course, from ${\cal H}_2$ into ${\cal H}_1$ at the moment $\tau+1$, and so
on.

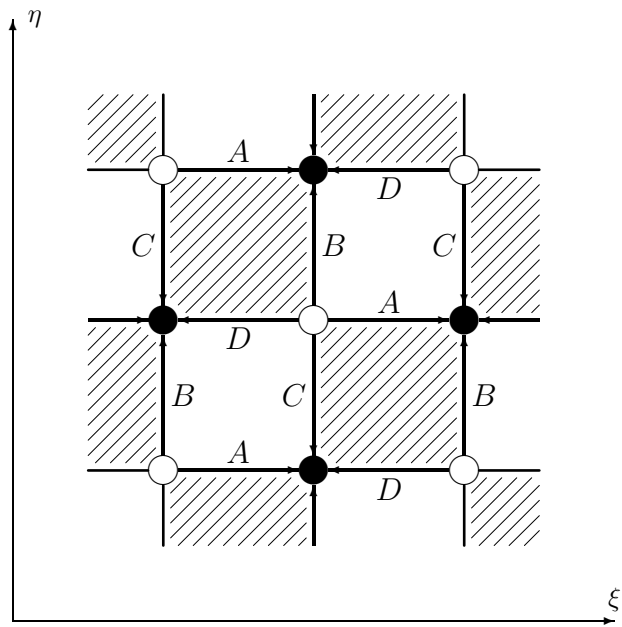
\begin{figure}
\begin{center}
\unitlength=1.00mm
\special{em:linewidth 1.00pt}
\linethickness{1.00pt}
\begin{picture}(80.00,80.00)
\put(20.00,20.00){\circle{4.00}}
\put(40.00,20.00){\circle*{4.00}}
\put(60.00,20.00){\circle{4.00}}
\put(60.00,40.00){\circle*{4.00}}
\put(40.00,40.00){\circle{4.00}}
\put(20.00,40.00){\circle*{4.00}}
\put(20.00,60.00){\circle{4.00}}
\put(40.00,60.00){\circle*{4.00}}
\put(60.00,60.00){\circle{4.00}}
\put(22.00,20.00){\vector(1,0){16.00}}
\put(20.00,22.00){\vector(0,1){16.00}}
\put(40.00,38.00){\vector(0,-1){16.00}}
\put(38.00,40.00){\vector(-1,0){16.00}}
\put(40.00,42.00){\vector(0,1){16.00}}
\put(42.00,40.00){\vector(1,0){16.00}}
\put(58.00,20.00){\vector(-1,0){16.00}}
\put(60.00,22.00){\vector(0,1){16.00}}
\put(60.00,58.00){\vector(0,-1){16.00}}
\put(58.00,60.00){\vector(-1,0){16.00}}
\put(22.00,60.00){\vector(1,0){16.00}}
\put(20.00,58.00){\vector(0,-1){16.00}}
\put(10.00,40.00){\vector(1,0){8.00}}
\put(40.00,70.00){\vector(0,-1){8.00}}
\put(70.00,40.00){\vector(-1,0){8.00}}
\put(40.00,10.00){\vector(0,1){8.00}}
\emline{20.00}{18.00}{1}{20.00}{10.00}{2}
\emline{18.00}{20.00}{3}{10.00}{20.00}{4}
\emline{10.00}{60.00}{5}{18.00}{60.00}{6}
\emline{20.00}{62.00}{7}{20.00}{70.00}{8}
\emline{60.00}{70.00}{9}{60.00}{62.00}{10}
\emline{62.00}{60.00}{11}{70.00}{60.00}{12}
\emline{70.00}{20.00}{13}{62.00}{20.00}{14}
\emline{60.00}{18.00}{15}{60.00}{10.00}{16}
\linethickness{0.40pt}
\put(0.00,0.00){\vector(1,0){80.00}}
\put(0.00,0.00){\vector(0,1){80.00}}
\put(30.00,61.00){\makebox(0,0)[cb]{\large$A$}}
\put(50.00,59.00){\makebox(0,0)[ct]{\large$D$}}
\put(19.00,50.00){\makebox(0,0)[rc]{\large$C$}}
\put(59.00,50.00){\makebox(0,0)[rc]{\large$C$}}
\put(30.00,39.00){\makebox(0,0)[ct]{\large$D$}}
\put(50.00,41.00){\makebox(0,0)[cb]{\large$A$}}
\put(61.00,30.00){\makebox(0,0)[lc]{\large$B$}}
\put(30.00,21.00){\makebox(0,0)[cb]{\large$A$}}
\put(50.00,19.00){\makebox(0,0)[ct]{\large$D$}}
\put(2.00,80.00){\makebox(0,0)[lc]{$\eta$}}
\put(80.00,2.00){\makebox(0,0)[cb]{$\xi$}}
\put(39.00,30.00){\makebox(0,0)[rc]{\large$C$}}
\put(21.00,30.00){\makebox(0,0)[lc]{\large$B$}}
\put(41.00,50.00){\makebox(0,0)[lc]{\large$B$}}
\special{em:linewidth 0.10pt}
\emline{23.00}{59.00}{17}{21.00}{57.00}{18}
\emline{21.00}{55.00}{19}{25.00}{59.00}{20}
\emline{27.00}{59.00}{21}{21.00}{53.00}{22}
\emline{29.00}{59.00}{23}{21.00}{51.00}{24}
\emline{21.00}{49.00}{25}{31.00}{59.00}{26}
\emline{33.00}{59.00}{27}{21.00}{47.00}{28}
\emline{21.00}{45.00}{29}{35.00}{59.00}{30}
\emline{37.00}{59.00}{31}{21.00}{43.00}{32}
\emline{22.00}{42.00}{33}{38.00}{58.00}{34}
\emline{39.00}{57.00}{35}{23.00}{41.00}{36}
\emline{25.00}{41.00}{37}{39.00}{55.00}{38}
\emline{39.00}{53.00}{39}{27.00}{41.00}{40}
\emline{29.00}{41.00}{41}{39.00}{51.00}{42}
\emline{39.00}{49.00}{43}{31.00}{41.00}{44}
\emline{33.00}{41.00}{45}{39.00}{47.00}{46}
\emline{39.00}{45.00}{47}{35.00}{41.00}{48}
\emline{37.00}{41.00}{49}{39.00}{43.00}{50}
\emline{43.00}{39.00}{51}{41.00}{37.00}{52}
\emline{41.00}{35.00}{53}{45.00}{39.00}{54}
\emline{47.00}{39.00}{55}{41.00}{33.00}{56}
\emline{49.00}{39.00}{57}{41.00}{31.00}{58}
\emline{41.00}{29.00}{59}{51.00}{39.00}{60}
\emline{53.00}{39.00}{61}{41.00}{27.00}{62}
\emline{41.00}{25.00}{63}{55.00}{39.00}{64}
\emline{57.00}{39.00}{65}{41.00}{23.00}{66}
\emline{42.00}{22.00}{67}{58.00}{38.00}{68}
\emline{59.00}{37.00}{69}{43.00}{21.00}{70}
\emline{45.00}{21.00}{71}{59.00}{35.00}{72}
\emline{59.00}{33.00}{73}{47.00}{21.00}{74}
\emline{49.00}{21.00}{75}{59.00}{31.00}{76}
\emline{59.00}{29.00}{77}{51.00}{21.00}{78}
\emline{53.00}{21.00}{79}{59.00}{27.00}{80}
\emline{59.00}{25.00}{81}{55.00}{21.00}{82}
\emline{57.00}{21.00}{83}{59.00}{23.00}{84}
\emline{17.00}{61.00}{85}{19.00}{63.00}{86}
\emline{19.00}{65.00}{87}{15.00}{61.00}{88}
\emline{13.00}{61.00}{89}{19.00}{67.00}{90}
\emline{19.00}{69.00}{91}{11.00}{61.00}{92}
\emline{10.00}{62.00}{93}{18.00}{70.00}{94}
\emline{16.00}{70.00}{95}{10.00}{64.00}{96}
\emline{10.00}{66.00}{97}{14.00}{70.00}{98}
\emline{12.00}{70.00}{99}{10.00}{68.00}{100}
\emline{68.00}{10.00}{101}{70.00}{12.00}{102}
\emline{70.00}{14.00}{103}{66.00}{10.00}{104}
\emline{64.00}{10.00}{105}{70.00}{16.00}{106}
\emline{70.00}{18.00}{107}{62.00}{10.00}{108}
\emline{61.00}{11.00}{109}{69.00}{19.00}{110}
\emline{67.00}{19.00}{111}{61.00}{13.00}{112}
\emline{61.00}{15.00}{113}{65.00}{19.00}{114}
\emline{63.00}{19.00}{115}{61.00}{17.00}{116}
\emline{42.00}{62.00}{117}{50.00}{70.00}{118}
\emline{48.00}{70.00}{119}{41.00}{63.00}{120}
\emline{41.00}{65.00}{121}{46.00}{70.00}{122}
\emline{44.00}{70.00}{123}{41.00}{67.00}{124}
\emline{41.00}{69.00}{125}{42.00}{70.00}{126}
\emline{52.00}{70.00}{127}{43.00}{61.00}{128}
\emline{45.00}{61.00}{129}{54.00}{70.00}{130}
\emline{56.00}{70.00}{131}{47.00}{61.00}{132}
\emline{49.00}{61.00}{133}{58.00}{70.00}{134}
\emline{59.00}{69.00}{135}{51.00}{61.00}{136}
\emline{53.00}{61.00}{137}{59.00}{67.00}{138}
\emline{59.00}{65.00}{139}{55.00}{61.00}{140}
\emline{57.00}{61.00}{141}{59.00}{63.00}{142}
\emline{39.00}{11.00}{143}{38.00}{10.00}{144}
\emline{39.00}{13.00}{145}{36.00}{10.00}{146}
\emline{34.00}{10.00}{147}{39.00}{15.00}{148}
\emline{39.00}{17.00}{149}{32.00}{10.00}{150}
\emline{30.00}{10.00}{151}{38.00}{18.00}{152}
\emline{37.00}{19.00}{153}{28.00}{10.00}{154}
\emline{26.00}{10.00}{155}{35.00}{19.00}{156}
\emline{33.00}{19.00}{157}{24.00}{10.00}{158}
\emline{22.00}{10.00}{159}{31.00}{19.00}{160}
\emline{29.00}{19.00}{161}{21.00}{11.00}{162}
\emline{21.00}{13.00}{163}{27.00}{19.00}{164}
\emline{25.00}{19.00}{165}{21.00}{15.00}{166}
\emline{21.00}{17.00}{167}{23.00}{19.00}{168}
\emline{62.00}{42.00}{169}{70.00}{50.00}{170}
\emline{70.00}{48.00}{171}{63.00}{41.00}{172}
\emline{65.00}{41.00}{173}{70.00}{46.00}{174}
\emline{70.00}{44.00}{175}{67.00}{41.00}{176}
\emline{69.00}{41.00}{177}{70.00}{42.00}{178}
\emline{61.00}{43.00}{179}{70.00}{52.00}{180}
\emline{70.00}{54.00}{181}{61.00}{45.00}{182}
\emline{61.00}{47.00}{183}{70.00}{56.00}{184}
\emline{70.00}{58.00}{185}{61.00}{49.00}{186}
\emline{61.00}{51.00}{187}{69.00}{59.00}{188}
\emline{67.00}{59.00}{189}{61.00}{53.00}{190}
\emline{61.00}{55.00}{191}{65.00}{59.00}{192}
\emline{63.00}{59.00}{193}{61.00}{57.00}{194}
\emline{18.00}{38.00}{195}{10.00}{30.00}{196}
\emline{10.00}{32.00}{197}{17.00}{39.00}{198}
\emline{15.00}{39.00}{199}{10.00}{34.00}{200}
\emline{10.00}{36.00}{201}{13.00}{39.00}{202}
\emline{11.00}{39.00}{203}{10.00}{38.00}{204}
\emline{10.00}{28.00}{205}{19.00}{37.00}{206}
\emline{19.00}{35.00}{207}{10.00}{26.00}{208}
\emline{10.00}{24.00}{209}{19.00}{33.00}{210}
\emline{19.00}{31.00}{211}{10.00}{22.00}{212}
\emline{11.00}{21.00}{213}{19.00}{29.00}{214}
\emline{19.00}{27.00}{215}{13.00}{21.00}{216}
\emline{15.00}{21.00}{217}{19.00}{25.00}{218}
\emline{19.00}{23.00}{219}{17.00}{21.00}{220}
\end{picture}
\end{center}
\caption{Integrable dynamical system in the 2+1-dimensional space-time}
\label{Fig1}
\end{figure}

Let each of the spaces ${\cal H}_1$, ${\cal H}_2$ be a direct sum of $lm/2$
identical subspaces of dimension $d$, where $l,m$ are even numbers. Let us
imagine these subspaces as situated
at the vertices of the square lattice on the torus
of the sizes $l\times m$ (which will mean the periodic boundary conditions in
both discrete space variables). Let the subspaces be arranged in checkerboard
fashion, as in Fig.~\ref{Fig1}, where the empty circles correspond to
subspaces of the space ${\cal H}_1$, while the filled circles correspond to
those of the space ${\cal H}_2$.

Let then the operators $A,\ldots D$ be such that the image of each of the
mentioned $d$-dimensional subspaces with respect to, say, operator $A$ lies in
the $d$-dimensional subspace of ${\cal H}_2$ at which points the arrow marked
``$A$'' that links these two subspaces (Fig.~\ref{Fig1}). Analogously, the
restrictions on $B$, $C$, $D$ are depicted in Fig.~\ref{Fig1} (see also
formula~(\ref{3.8}) for non-degenerate $A,\ldots D$). Thus, to each link of
the lattice a $d\times d$ matrix is attached that is a block of one of the
``large'' matrices $A,\ldots D$. Let us shade half of the squares of the
lattice in a checkerboard way, as in Fig.~\ref{Fig1}. One can verify that the
evolution of the system may be described as follows.

At the first step, each of the four $d\times d$ matrices that correspond to
the arrows surrounding each shaded square is transformed into a matrix
expressed through just these four matrices. This goes according to the
following formulae, in which the $d\times d$ blocks are somewhat freely
denoted by the same letters $A,\ldots D$ as the ``large'' matrices:
\be A\longrightarrow(A-BD^{-1}C)^{-1},\label{3.1a}\ee
\be B\longrightarrow(B-AC^{-1}D)^{-1},\label{3.1b}\ee
\be C\longrightarrow(C-DB^{-1}A)^{-1},\label{3.1c}\ee
\be D\longrightarrow(D-CA^{-1}B)^{-1}.\label{3.1d}\ee
However, the formulae~(\ref{3.1a}--\ref{3.1d}) apply equally to the ``large''
matrices.

After the transformation~(\ref{3.1a}--\ref{3.1d}), all the arrows reverse, and
at the second step the non-shaded squares are engaged in the same way
according to
the same formulae~(\ref{3.1a}--\ref{3.1d}). Then everything is repeated. Thus,
the evolution is of hyperbolic nature: each local perturbation spreads not
faster than one unit of length per unit of time.

Let us clarify the symmetries of vacuum curves and divisors $D_X$ in this
``reduced'' model. Let us introduce two integer-valued coordinates $\xi,\eta$
for the vertices of the lattice, so that $\xi$ increases by 1 in passing from
a vertex one step to the right, and $\eta$ increases by 1 in passing one step
upwards. $\xi$ and $\eta$ are defined modulo $l$ and $m$ respectively. A
$d$-dimentional subspace of ${\cal H}_1$ or ${\cal H}_2$ will be denoted
${\cal H}_{\xi\eta}$ if it corresponds to a vertex with coordinates
$\xi,\eta$. Consider a linear transformation in spaces ${\cal H}_1$ and
${\cal H}_2$ consisting in multiplying the vectors of each subspace
${\cal H}_{\xi\eta}$ by $\omega^\xi_1,\quad$ $\omega_1$ being a fixed
primitive root of the $l$-th degree of unity:
$$\omega_1^l=1.$$
This corresponds to the following transformation of the operators
$A,\ldots D$ (from now on we speak of each of these operators ``as a whole'',
not of their blocks):
\be A\rightarrow\omega_1A,\;B\rightarrow B,\;C\rightarrow C,\;D\rightarrow
\omega_1^{-1}D.\label{3.2}\ee
Consider also another linear transformation in ${\cal H}_1$ and ${\cal H}_2$,
consisting in multiplying the vectors of each subspace ${\cal H}_{\xi\eta}$
by $\omega_2^\eta,\;$ $\omega_2$ being a fixed primitive root of the $m$-th
degree of unity:
$$\omega_2^m=1.$$
This corresponds to the following transformation:
\be A\rightarrow A,\;B\rightarrow\omega_2B,\;C\rightarrow\omega_2^{-1}C,\;
D\rightarrow D.\label{3.3}\ee

The vacuum curve of the operator $L$, which is given by equation~(\ref{2.4})
$$P(u,v)=\det(uA+B-uvC-vD)=0,$$
must be invariant under the transformations~(\ref{3.2}), (\ref{3.3}). This
leads to the invariance of the polynomial $P(u,v)$ with respect to the
following transformations $g_1$ and $g_2$:
\be g_1(u,v)=(\omega_1u,\omega_1^{-1}v),\label{3.4}\ee
\be g_2(u,v)=(\omega_2^{-1}u,\omega_2^{-1}v).\label{3.5}\ee
This invariance, then, leads to the following statement: only those
coefficients $a_{jk}$ are non-zero in the vacuum curve equation
(see~(\ref{2.2}), (\ref{2.3})) for the ``reduced'' model, for which
\be\left. \ba{l}j-k\equiv0({\rm mod}\;l),\smallskip\\j+k\equiv0({\rm mod}\;m).
\ea\right\}\label{3.6}\ee

As for the divisor $D_X$, let us recall that it consists of such points in the
curve $\Gamma$ in which vanishes the last coordinate of the vector $X$ (see
\cite{Krichever}), the latter being an eigenvector of the matrix~(\ref{2.6})
with zero eigenvalue:
\be(uA+B-uvC-vD)X(u,v)=0.\label{3.7}\ee
This immediately leads to the conclusion: the divisor $D_X$ is invariant with
respect to the transformations~(\ref{3.4},~\ref{3.5}).

Under some additional condition, the inverse statement also holds: if the
curve $\Gamma$ and divisor $D_X$ are invariant under the
transformations~(\ref{3.4}), (\ref{3.5}), then the corresponding $L$-operator
comes from a ``reduced'' model described in this section. For instance, this
is true if $l/2$ and $m/2$ are relatively prime numbers. If these numbers are
not relatively prime, some conditions are to be imposed on the divisor $D_X$.
To avoid going into details of this latter case, let us not consider it here.

Thus, let an operator $L=\left(\!\ba{cc}A&B\\C&D\ea\!\right)$ be given, $A,
\ldots
D$ being $n\times n$ matrices, $n=(lm/2)d,\quad l$ and $m$ even, and $l/2$ and
$m/2$ being relatively prime. Let the vacuum curve $\Gamma$ of the operator
$L$ and the divisor $D_X$ be invariant under the action of the group $\cal G$
generated by its elements $g_1,g_2$~(\ref{3.4},~\ref{3.5}), $\;\omega_1$ and
$\omega_2$ being primitive roots of degrees $l$ and $m$ of unity. Then the
linear space in which operators $A,\ldots D$ act decomposes into a direct sum
of $lm/2\;\;$ $d$-dimensional subspaces ${\cal H}_{\xi\eta},\;\;$ $\xi$ and
$\eta$ being integers modulo $l$ and $m$ respectively and such that $\xi+\eta$
is an even number, and the following equalities between the images of these
subspaces hold (in a ``generic'' case of non-degenerate $A,\ldots D$):
\be A{\cal H}_{\xi-1,\eta+1}=B{\cal H}_{\xi\eta}=C{\cal H}_{\xi,\eta+2}=
D{\cal H}_{\xi+1,\eta+1}.\label{3.8}\ee
The equalities~(\ref{3.8}) mean exactly that one is in the situation of
Fig.~\ref{Fig1}.

Let us prove the above statements. First, the natural projection from the
curve $\Gamma$ to its factor $\Gamma/\cal G$ has no branch points (here the
fact that $l/2$ and $m/2$ are relatively prime is used to demonstrate that
ramification does not occur when $u$ or $v$ equals zero or infinity). Thus,
the $n$-dimensional linear space of meromorphic functions $x(z)=x(u,v)$ whose
pole divisor is $D_X$ decomposes into a direct sum of subspaces of equal
dimensions corresponding to the characters of (commutative) group $\cal G$.
Each of these subspaces consists of functions $x(z)$ satisfying relations
$$x(gz)=\chi_{\xi\eta}(g)x(z),$$
the character $\chi_{\xi\eta}$ being a scalar factor
$$\chi_{\xi\eta}(g)=\omega_1^{\xi a}\omega_2^{\eta b},$$
where
$$g=g_1^ag_2^b.$$
The equality $g_1^{l/2}g_2^{m/2}=1$ means that $\xi+\eta$ must be an even
number.

The components of the vector $X(z)$ are exactly the functions $x(z)$. In an
appropriate basis, $d$ components correspond to each character
$\chi_{\xi\eta}$. Let us denote ${\cal H}_{\xi\eta}$ the set of vectors with
other components equal to zero. Now, the equalities~(\ref{3.8}) are to be
proved to end this section.

Consider the decomposition of vector $X(u,v)$ into a sum
$$X(u,v)=\sum_{\xi,\eta}X_{\xi,\eta}(u,v),$$
where $X_{\xi,\eta}\in{\cal H}_{\xi,\eta}$. Then
$$X_{\xi,\eta}\bigl( g(u,v)\bigr)=\chi_{\xi,\eta}(g)X_{\xi,\eta}(u,v).$$
Consider the sum
\be \sum_{g\in{\cal G}}\chi_{\xi,\eta}(g^{-1})g\{(uA+B-uvC-vD)X(u,v)\}=0
\label{3.9}\ee
(which is equal to zero because of~(\ref{3.7})). The action of $g$ upon the
braces in~(\ref{3.9}) means that each $u$ and $v$ in the braces is
transformed according to~(\ref{3.4}), (\ref{3.5}), i.e. $u$ changes into
$\chi_{1,-1}(g)u$, and $v$ changes into $\chi_{-1,-1}(g)v$. The
equality~(\ref{3.9}) gives thus
\be uAX_{\xi-1,\eta+1}(u,v)+BX_{\xi,\eta}(u,v)-uvCX_{\xi,\eta+2}(u,v)-
vX_{\xi+1,\eta+1}(u,v)D=0.\label{3.10}\ee
Let us set $u=0$ in~(\ref{3.10}). Then $v$ can take $n$ different values $v_j$
satisfying relation $P(0,v_j)=0$. To these values $v_j$ correspond $d$
linearly independent vectors $X_{\xi\eta}(0,v_j)$, and also $d$ vectors
$X_{\xi+1,\eta+1}(0,v_j)$. Thus, the equalities
$$BX_{\xi\eta}(0,v_j)=v_jDX_{\xi+1,\eta+1}(0,v_j)$$
that result from~(\ref{3.10}) give
$$B{\cal H}_{\xi\eta}=D{\cal H}_{\xi+1,\eta+1}.$$
Analogously, one can as well obtain the rest of equalities~(\ref{3.8}).

\section{Connection to dimer model}\label{dimers}

As has been demonstrated, the integrals of motion of the dynamical system
of Section~\ref{gensys} and its reductions (if the {\em even} degrees
of the transformation~(\ref{1.1}) are considered)
are the coefficients $a_{jk}$ of the vacuum curve~(\ref{2.3}). These
coefficients are determined up to a common factor, so they may be divided by
$a_{00}$. As one can see, the resulting coefficients are those of the
polynomial
\be\det\bigl(1+uAB^{-1}-vDB^{-1}-uvCB^{-1}\bigr).\label{determ}\ee
In other words, the determinant~(\ref{determ}) is an integral of motion for
any $u,v$.

\begin{figure}
\begin{center}
\unitlength=1.00mm
\special{em:linewidth 0.4pt}
\linethickness{0.4pt}
\begin{picture}(57.00,42.00)
\put(21.00,6.00){\circle{2.00}}
\put(6.00,6.00){\circle*{2.00}}
\put(6.00,21.00){\circle{2.00}}
\put(36.00,21.00){\circle{2.00}}
\put(21.00,21.00){\circle*{2.00}}
\put(36.00,6.00){\circle*{2.00}}
\put(21.00,36.00){\circle{2.00}}
\put(6.00,36.00){\circle*{2.00}}
\put(36.00,36.00){\circle*{2.00}}
\put(51.00,21.00){\circle*{2.00}}
\put(51.00,6.00){\circle{2.00}}
\put(51.00,36.00){\circle{2.00}}
\put(20.00,6.00){\vector(-1,0){13.00}}
\put(22.00,6.00){\vector(1,0){13.00}}
\put(50.00,6.00){\vector(-1,0){13.00}}
\put(21.00,20.00){\vector(0,-1){13.00}}
\put(6.00,20.00){\vector(0,-1){13.00}}
\put(36.00,20.00){\vector(0,-1){13.00}}
\put(51.00,20.00){\vector(0,-1){13.00}}
\put(6.00,35.00){\vector(0,-1){13.00}}
\put(21.00,35.00){\vector(0,-1){13.00}}
\put(51.00,35.00){\vector(0,-1){13.00}}
\put(7.00,21.00){\vector(1,0){13.00}}
\put(35.00,21.00){\vector(-1,0){13.00}}
\put(37.00,21.00){\vector(1,0){13.00}}
\put(50.00,36.00){\vector(-1,0){13.00}}
\put(22.00,36.00){\vector(1,0){13.00}}
\put(20.00,36.00){\vector(-1,0){13.00}}
\put(14.00,37.00){\makebox(0,0)[cb]{$-vd$}}
\put(28.00,37.00){\makebox(0,0)[cb]{$ua$}}
\put(44.00,37.00){\makebox(0,0)[cb]{$-vd$}}
\put(14.00,22.00){\makebox(0,0)[cb]{$ua$}}
\put(28.00,22.00){\makebox(0,0)[cb]{$-vd$}}
\put(44.00,22.00){\makebox(0,0)[cb]{$ua$}}
\put(44.00,7.00){\makebox(0,0)[cb]{$-vd$}}
\put(28.00,7.00){\makebox(0,0)[cb]{$ua$}}
\put(14.00,7.00){\makebox(0,0)[cb]{$-vd$}}
\put(7.00,29.00){\makebox(0,0)[lc]{$b^{-1}$}}
\put(22.00,29.00){\makebox(0,0)[lc]{$-uvc$}}
\put(37.00,29.00){\makebox(0,0)[lc]{$b^{-1}$}}
\put(52.00,29.00){\makebox(0,0)[lc]{$-uvc$}}
\put(52.00,14.00){\makebox(0,0)[lc]{$b^{-1}$}}
\put(37.00,14.00){\makebox(0,0)[lc]{$-uvc$}}
\put(22.00,14.00){\makebox(0,0)[lc]{$b^{-1}$}}
\put(7.00,14.00){\makebox(0,0)[lc]{$-uvc$}}
\put(6.00,42.00){\vector(0,-1){5.00}}
\put(21.00,42.00){\vector(0,-1){5.00}}
\put(36.00,42.00){\vector(0,-1){5.00}}
\put(51.00,42.00){\vector(0,-1){5.00}}
\put(0.00,36.00){\vector(1,0){5.00}}
\put(0.00,6.00){\vector(1,0){5.00}}
\put(57.00,21.00){\vector(-1,0){5.00}}
\emline{0.00}{21.00}{1}{5.00}{21.00}{2}
\emline{6.00}{5.00}{3}{6.00}{0.00}{4}
\emline{21.00}{0.00}{5}{21.00}{5.00}{6}
\emline{36.00}{5.00}{7}{36.00}{0.00}{8}
\emline{51.00}{0.00}{9}{51.00}{5.00}{10}
\emline{52.00}{6.00}{11}{57.00}{6.00}{12}
\emline{57.00}{36.00}{13}{52.00}{36.00}{14}
\put(36.00,35.00){\vector(0,-1){13.00}}
\end{picture}
\caption{The ways along these arrows are connected with both the vacuum curve
and the dimer model}
\label{arrows}
\end{center}
\end{figure}
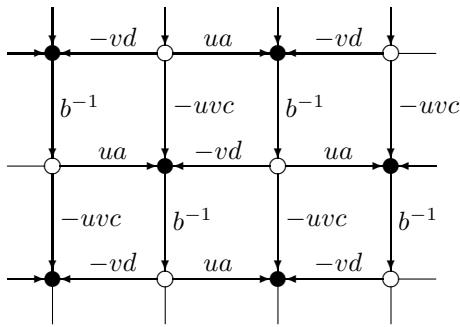

Let us turn now to the model from section~\ref{reduction}, that is to the
model in 2+1-dimensional discrete space-time with periodic boundary
conditions, and let the dimension $d$ of the linear space corresponding to
each vertex be equal to 1. Each of the ``small'' matrices $A,B,C,D$
corresponding to the links will then be a single (depending on the link)
number $a,b,c$ or $d$. It is well known that the determinant of any
$N\times N$ matrix is a sum of its matrix elements products corresponding in
a certain way to the permutations of $N$ objects, while each permutation
decomposes into a product of the cyclic ones. In our situation, the cyclic
permutations correspond to the non-selfintersecting closed paths (contours)
going along the arrows of the following diagram~(Fig.~\ref{arrows}) (thus,
general permutations correspond to the sets of non-intersecting paths).
To each closed path corresponds the product of the weights $ua,-uvc,-vd,
b^{-1}$ on its links, and, to get right signs for the terms of which the
determinant~(\ref{determ}) is made up, one should add a minus sign to each
such product containing an {\em even} number of the factors $b^{-1}$.

\begin{remark} \label{r1}
Another way to obtain right signs is: to multiply each $b$ by $-1$ and then
multiply each product corresponding to a closed path (and containing
{\em any} number of $b$'s) by $-1$.
\end{remark}

\begin{figure}
\begin{center}
\unitlength=1.00mm
\special{em:linewidth 0.2pt}
\linethickness{0.2pt}
\begin{picture}(57.00,42.00)
\put(21.00,6.00){\circle{2.00}}
\put(6.00,21.00){\circle{2.00}}
\put(36.00,21.00){\circle{2.00}}
\put(21.00,21.00){\circle*{2.00}}
\put(36.00,6.00){\circle*{2.00}}
\put(21.00,36.00){\circle{2.00}}
\put(6.00,36.00){\circle*{2.00}}
\put(36.00,36.00){\circle*{2.00}}
\put(51.00,21.00){\circle*{2.00}}
\put(51.00,6.00){\circle{2.00}}
\put(51.00,36.00){\circle{2.00}}
\put(6.00,6.00){\circle*{2.00}}
\emline{7.00}{36.00}{1}{20.00}{36.00}{2}
\emline{22.00}{36.00}{3}{35.00}{36.00}{4}
\emline{37.00}{36.00}{5}{50.00}{36.00}{6}
\emline{52.00}{36.00}{7}{57.00}{36.00}{8}
\emline{0.00}{36.00}{9}{5.00}{36.00}{10}
\emline{5.00}{21.00}{11}{0.00}{21.00}{12}
\emline{0.00}{6.00}{13}{5.00}{6.00}{14}
\emline{7.00}{6.00}{15}{20.00}{6.00}{16}
\emline{22.00}{6.00}{17}{35.00}{6.00}{18}
\emline{37.00}{6.00}{19}{50.00}{6.00}{20}
\emline{52.00}{6.00}{21}{57.00}{6.00}{22}
\emline{57.00}{21.00}{23}{52.00}{21.00}{24}
\emline{50.00}{21.00}{25}{37.00}{21.00}{26}
\emline{35.00}{21.00}{27}{22.00}{21.00}{28}
\emline{20.00}{21.00}{29}{7.00}{21.00}{30}
\emline{6.00}{37.00}{31}{6.00}{42.00}{32}
\emline{36.00}{42.00}{33}{36.00}{37.00}{34}
\emline{21.00}{35.00}{35}{21.00}{22.00}{36}
\emline{51.00}{22.00}{37}{51.00}{35.00}{38}
\emline{36.00}{20.00}{39}{36.00}{7.00}{40}
\emline{6.00}{7.00}{41}{6.00}{20.00}{42}
\emline{21.00}{5.00}{43}{21.00}{0.00}{44}
\emline{51.00}{0.00}{45}{51.00}{5.00}{46}
\special{em:linewidth 2pt}
\linethickness{2pt}
\emline{6.00}{5.00}{47}{6.00}{0.00}{48}
\emline{21.00}{7.00}{49}{21.00}{20.00}{50}
\emline{6.00}{22.00}{51}{6.00}{35.00}{52}
\emline{36.00}{35.00}{53}{36.00}{22.00}{54}
\emline{21.00}{37.00}{55}{21.00}{42.00}{56}
\emline{51.00}{42.00}{57}{51.00}{37.00}{58}
\emline{51.00}{20.00}{59}{51.00}{7.00}{60}
\emline{36.00}{5.00}{61}{36.00}{0.00}{62}
\end{picture}
\caption{The standard dimer configuration}
\label{standard}
\end{center}
\end{figure}
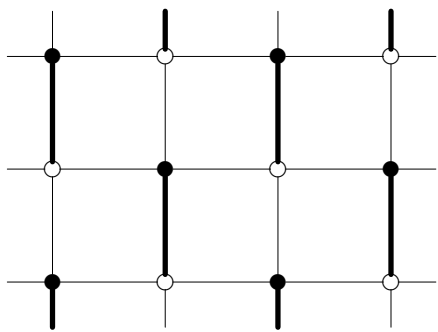

It turns out that the determinant~(\ref{determ}) is connected with the
statistical sum of the well known dimer model~\cite{dimersbook}.
Let us define the correspondence between the sets of paths and the dimer
configurations as follows. Let the empty set of paths correspond to the
``standard'' dimer configuration, the dimers being placed on the
``$B$-links'' (Fig.~\ref{standard}). For a non-empty set of paths, let us
change the standard configuration along all the paths, replacing each dimer by
a free link and vice versa. One can verify that this is a bijective
correspondence.

The statistical sum being considered, let the weights $-b$ (not $b^{-1}$)
correspond to the ``$B$-links'', while to the other links correspond
the unchanged weights $ua,-vd,-uvc$. Then one can see that the statistical
sum, if multiplied by $\prod\nolimits_{\mbox{\footnotesize over all links}}
(-b^{-1})$
(let us call the result the normalized statistical sum), consists of the same
terms as the determinant~(\ref{determ}), up to different signs of some of
them. Let us emphasize that the dimer model is, of course, {\em
inhomogeneous}: the weights $a,b,c,d$ are different for different links.

Let us study these signs in detail. Note that the conditions of
non-inter\-secting and non-selfintersecting impose strong restrictions on the
possible path configurations. Every closed path on the torus is
homologically equivalent to a linear combination with integer coefficients of
two {\em basis cycles} $\bf a$ and $\bf b$ whose intersection number is~1
(I use the boldface font for cycles, because the letters $a,b\ldots$ are
already in use). If the torus is cut along a closed non-selfintersecting path
$\bf c$ not equivalent to zero, the result will be homeomorphic to the lateral
surface of a cylinder (this follows, e.g., from~\cite{rDNF}, chapter~1,
section~3). Then the contour~$\bf d$ going along a generatrix of the cylinder
in a properly chosen direction has the intersection number~1 with the
contour~$\bf c$. The intersection number being bilinear and
integer-valued, we find that if the contour~$\bf c$ is
homologically equivalent to a sum $l{\bf a}+m{\bf b}$, then $l$ and $m$
cannot have common divisors (not equal to~$\pm1$). Thus, the following lemma
is valid.

\begin{lemma} \label{l2}
Every closed non-selfintersecting path on the torus is homologically
equivalent to a linear combination of the basis paths $\bf a$ and $\bf b$
with relatively prime integer coefficients.
\end{lemma}

Now let us pass to the case of several contours on the torus. If they do not
intersect, their intersection numbers equal~0 (of course) and thus their
homological classes must be proportional to one another. This together with
Lemma~\ref{l2} leads to the following lemma.

\begin{lemma} \label{l1}
Several closed non-intersecting and non-selfintersecting paths going along
the arrows on the torus, as in Fig.~\ref{arrows}, are necessarily all
homologically equivalent to one another.
\end{lemma}

If two paths are homologically equivalent, then the terms of the same degrees
in $u$ and $v$ correspond to them (one can see in Fig.~\ref{arrows} that the
different ways round an ``elementary square'' yield the same degrees of
$u$ and $v$). Let the basis paths $\bf a$ and $\bf b$ yield the terms
proportional to $x=u^{\alpha_1}v^{\beta_1},y=u^{\alpha_2}v^{\beta_2}$
correspondingly (with the factors of proportionality not depending on
$u,v$). According to Lemma~\ref{l2}, the determinant~(\ref{determ}) and the
statistical sum of the dimer model are polynomials in $x,y$. The following
lemma sums up this section.

\begin{lemma} \label{l3}
Let $f(x,y)$ and $s(x,y)$ be the determinant~(\ref{determ}) and the
normalized statistical sum of the dimer model considered as functions of
$x$ and $y$. Then
\be s(x,y)=\frac{1}{2}\bigl(-f(x,y)+f(-x,y)+f(x,-y)+f(-x,-y)\bigr),\label{fs}
\ee
\be f(x,y)=\frac{1}{2}\bigl(-s(x,y)+s(-x,y)+s(x,-y)+s(-x,-y)\bigr).\label{sf}
\ee
\end{lemma}

{\it Proof.} If the normalized statistical sum consists of the terms
$$c_{jk}x^jy^k=c_{jk}(u^{\alpha_1}v^{\beta_1})^j(u^{\alpha_2}v^{\beta_2})^k,$$
then the determinant consists of the same terms multiplied by
$$(-1)^{\mbox{\footnotesize number of contours}}=(-1)^{\mbox{\footnotesize
g.c.d.}(j,k)}=(-1)^{jk+j+k}$$
(here Remark~\ref{r1} and Lemmas~\ref{l2} and~\ref{l1} are used). This means
that the signs of all the terms must be changed except where both numbers
$j$ and $k$ are even. This is exactly what the formulae~(\ref{fs},~\ref{sf})
do. The lemma is proved.

\section{The discrete analog of Lax pair and a generalization of the dynamical
system}

Now let us return from the reduction of Section~\ref{reduction} to general
matrices $L=\left(\ba{cc}A&B\\C&D\ea\right)$. Let us consider the evolution
described in Section~\ref{gensys} from another viewpoint. Denote
$$(L^{-1})^t=\left(\ba{cc}\tilde A&\tilde B\\\tilde C&\tilde D\ea\right).$$
This means that
\be\left(\ba{cc}\tilde A&\tilde C\\\tilde B&\tilde D\ea\right)
\left(\ba{cc}A&B\\C&D\ea\right)=\mbox{\large\bf1}.\label{4.1}\ee
It follows from the equality~(\ref{4.1}) that
\begin{eqnarray}\tilde AA+\tilde CC&=&\tilde BB+\tilde DD,\label{4.2}\\
\tilde AB+\tilde CD&=&0,\nonumber\\
\tilde BA+\tilde DC&=&0.\nonumber\end{eqnarray}
These three equations are equivalent to the fact that the following equality
holds for any complex $u$:
\be-(\tilde A-u\tilde B)^{-1}(\tilde C-u\tilde D)=(uA+B)(uC+D)^{-1}.
\label{4.3}\ee
Vice versa, from~(\ref{4.3}) follows
$$\tilde L^tL=\left(\ba{cc}F&0\\0&F\ea\right),$$
$F$ being equal to both sides of~(\ref{4.2}), i.e.
$$\tilde L=\left(\ba{cc}F&0\\0&F\ea\right)(L^{-1})^t.$$
It is clear that with any choice of $F$ the matrix $\tilde L$ belongs to the
same equivalence class. The formula~(\ref{4.3}) defines the same evolution
in the space of these classes as it was in Section~\ref{gensys}, with the
agreement that the operators without a tilde correspond to the moment of time
$\tau$, while those with a tilde correspond to the moment $\tau+1$.

The formula~(\ref{4.3}) suggests the following generalization. Let, from now
on, $A(u)$ and $B(u)$ be matrices depending polynomially on $u$:
\be A(u)=A_0+A_1+\ldots+A_{m_A}u^{m_A},\label{4.4}\ee
\be B(u)=B_0+B_1+\ldots+B_{m_B}u^{m_B}.\label{4.5}\ee
We will look for matrices $\tilde A(u),\tilde B(u)$---the matrix polynomials
of the same degrees $m_A$ and $m_B$ in $u$---that satisfy, for any $u$, the
equation
\be\tilde B(u)^{-1}\tilde A(u)=A(u)B(u)^{-1}.\label{4.6}\ee
The relation~(\ref{4.6}) provides what is called a discrete analog of the Lax
$L,A$-pair, which means here that the operators $\tilde A(u)\tilde B(u)^{-1}$
and $A(u)B(u)^{-1}$ (which are playing the role of $L$ of the pair) are
``isospectral deformations'' of one another:
$$\tilde A(u)\tilde B(u)^{-1}=\tilde A(u)A(u)B(u)^{-1}\tilde A(u)^{-1}.$$

Let $v$ be an eigenvalue of both sides of~(\ref{4.6}). Let $Y(u,v)$ be the
corresponding eigenvector normalized, as in Section~\ref{vacuum}, so that its
last coordinate equals unity, and let $X(u,v)$ be the vector proportional to
$B(u)^{-1}Y(u,v)$ and normalized in the same way. One can verify that this may
be described by the following formula ($h(u,v)$ being a scalar factor):
\be\left(\ba{c}A(u)\\B(u)\ea\right)X(u,v)=h(u,v)\left(\ba{c}v\\1\ea\right)
\otimes Y(u,v),\label{4.7}\ee
which is in obvious analogy to~(\ref{2.5}). The divisor equivalence is
\be mD_u+D_X\sim D_v+D_Y,\label{4.8}\ee
$D_u$ and $D_v$ being pole divisors of the functions $u$ and $v,\quad m={\rm
max}(m_A,m_B).$

For a given $u$, the eigenvalues $v$ come from the equation
$$P(u,v)=\det\bigl(A(u)-vB(u)\bigr)=0.$$
It defines an algebraic curve $\Gamma$---``generalized vacuum curve''. Let us
calculate the genus $g$ of the curve $\Gamma$. First, we need to know the
number of branch points of the projection
\be(u,v)\longrightarrow u\label{4.9}\ee
of the curve $\Gamma$ onto the complex plane.

Consider $P(u,v)$ as a polynomial in $v$:
\be P(u,v)=a_0(u)+a_1(u)+\ldots+a_n(u)v^n.\label{4.10}\ee
One can verify that $a_j(u)$ has a degree
\be{\rm deg}\;a_j(u)=(n-j)m_A+j_B.\label{4.11}\ee
 From this one can deduce that the discriminant of $P(u,v)$ considered as a
polynomial in $v$ is a polynomial of degree
$$b=(m_A+m_B)n(n-1)$$
in $u$. The mapping~(\ref{4.9}) being $n$-sheeted and the number of branch
points equalling $b$, one obtains from the Riemann---Hurwitz formula that
\be g=(n-1)\left(\frac{m_A+m_B}{2}n-1\right).\label{4.12}\ee

So, the following construction has been described. Given two polynomial matrix
functions $A(u)$ and $B(u)$, one considers the meromorphic matrix function
$A(u)B(u)^{-1}$ (or else $B(u)^{-1}A(u)$), and from this function the
algebro-geometrical objects arise: the generalized vacuum curve $\Gamma$ and
the linear equivalence class of the pole divisor $D_Y$ (or, respectively,
$D_X$) of the eigenvectors of the mentioned meromorphic matrix function.
Instead of the pair $(A(u),B(u))$, it is sufficient to indicate its
equivalence class with respect to gauge transformations
\be A(u)\rightarrow GA(u)H,\;B(u)\rightarrow GB(u)H;\label{4.13}\ee
instead of the function $A(u)B(u)^{-1}$, its equivalence class with respect to
transformations
$$A(u)B(u)^{-1}\rightarrow GA(u)B(u)^{-1}G^{-1}$$
will suffice. Then it turns out that the correspondence between such
equivalence classes (either of the pairs $(A(u),B(u))$ or the functions
$A(u)B(u)^{-1}$) and the abovementioned algebro-geometrical objects is a
birational isomorphism, the divisors $D_X$ and $D_Y$ being of degree $g+n-1$,
as in Section~\ref{vacuum}.

The easiest way to show this is to start from a given curve $\Gamma$ defined
by the equation
$$P(u,v)=\sum_{j=0}^n\;\sum_{k=0}^{(n-j)m_A+jm_B}\;a_{jk}v^ju^k=0$$
(compare with~(\ref{4.10}, \ref{4.11})) and a divisor $D_X$ in it of degree
$g+n-1$. The number of coefficients $a_{jk}$ minus one common factor equals
\be(n+1)\left(\frac{m_A+m_B}{2}n+1\right)-1.\label{4.14}\ee
The linear equivalence class of divisor $D_X$ is defined, as is known, by $g$
parameters. Adding up the expressions~(\ref{4.14}) and~(\ref{4.12}), one gets
the total of
\be(m_A+m_B)n^2+1\label{4.15}\ee
parameters.

Then, the gauge equivalence class of the pair $(A(u),B(u))$ is constructed out
of relation~(\ref{4.7}). To give more details, one must at first choose a
divisor $D_Y$ satisfying the equivalence~(\ref{4.8}). Then the poles and zeros
of the function $h(u,v)$ are determined. For $X(u,v)$ and $Y(u,v)$ one must
take columns consisting each of $n$ linearly independent meromorphic functions
with corresponding pole divisors. The arbitrariness in these constructions
leads exactly to the fact that $A(u)$ and $B(u)$ are determined up to a
transformation~(\ref{4.13}).

The pair $(A(u),B(u))$, up to a scalar common factor, is determined by
$(m_A+m_B+2)n^2-1$ parameters (see~(\ref{4.4}, \ref{4.5})). In taking the
gauge equivalence class, the number of parameters is reduced by $2(n^2-1)$.
The result is again~(\ref{4.15}). This means that, indeed, to a generic pair
$(A(u),B(u))$ corresponds a divisor $D_X$ of degree $g+n-1$ and the
correspondence
$$\bigl(\mbox{gauge equivalence class of the pair }(A(u),B(u))
\bigr)\longleftrightarrow\bigl(\Gamma,\mbox{ class of }D_X\bigr)$$
is a birational isomorphism.

Now let us recall that $Y(u,v)$ was defined as an eigenvector of the operator
$A(u)B(u)^{-1}$, while $X(u,v)$, as is easily seen, is an eigenvector of
$B(u)^{-1}A(u)$. The relation~(\ref{4.6}) means that for the pair
$(\tilde A(u),\tilde B(u))$ its vector $\tilde X(u,v)$ is nothing else than
$Y(u,v)$, i.e.\ the equivalence holds
\be D_{\tilde X}\sim D_X+(mD_u-D_v).\label{4.16}\ee

Now, assuming that if a quantity without a tilde corresponds to the moment of
time $\tau$ then that with a tilde corresponds to $\tau+1$, one comes to a
conclusion that to the adding of unity to the time corresponds a constant
shift~(\ref{4.16}) in the Jacobian of the curve~$\Gamma$. Thus, the dynamics
of the system in this section, as well as in Section~\ref{vacuum}, linearizes.

\section{Discussion}

In this paper I study a dynamical system in discrete time, i.e.\ a mapping and
its iterations, acting on finite sets of $n\times n$ matrices. The system
appears in several modifications, on which depends the number of matrices as
well as the additional conditions that may be imposed on them. The ``law of
motion'' is formulated in a rather simple way, and a large number of
``integrals of motion'' turn out to exist and be the coefficients of the
``vacuum curve''---the object coming from the theory of the quantum
Yang---Baxter equation. If the motion in the system is considered up to a
``gauge transformation'', the system is integrable in the sense that there
exists a birational isomorphism between the ``phase space'' and the set of
pairs (a vacuum curve, an element of its Picard group), so that in the process
of ``motion'' the vacuum curve doesn't change, while the element of its Picard
group depends on the time linearly. Thus, the Cauchy problem is solved through
the following scheme: the initial point in the phase space $\longrightarrow$
the vacuum curve and the element of its Picard group at the initial moment of
time $\longrightarrow$ the same at the moment $\tau$ $\longrightarrow$ the
element of the phase space at the moment $\tau$.

Connections with statistical physics are exposed in Section~\ref{dimers},
where a special reduction of the model is considered. Note that there exists
one more dynamical system in 2+1-dimensional discrete space-time that is also
connected with statistical physics and seems to be completely integrable. This
system is as follows. Consider the inhomogeneous Ising model on the {\it
triangle} lattice. Imagine this lattice as consisting of triangles of the
form%
\unitlength=0.50mm
\special{em:linewidth 0.6pt}
\linethickness{0.2pt}
\begin{picture}(8.00,8.00)
\emline{0.00}{0.00}{1}{8.00}{0.00}{2}
\emline{8.00}{0.00}{3}{4.00}{8.00}{4}
\emline{4.00}{8.00}{5}{0.00}{0.00}{6}
\end{picture}
\ \ and perform for each of them  the ``triangle---star''
transformation~\cite{Baxterbook}:
\begin{center}
\unitlength=0.5mm
\special{em:linewidth 0.6pt}
\linethickness{0.2pt}
\begin{picture}(27.00,8.00)
\emline{0.00}{0.00}{1}{8.00}{0.00}{2}
\emline{8.00}{0.00}{3}{4.00}{8.00}{4}
\emline{4.00}{8.00}{5}{0.00}{0.00}{6}
\put(10.00,4.00){\vector(1,0){7.00}}
\emline{19.00}{0.00}{7}{23.00}{3.00}{8}
\emline{23.00}{3.00}{9}{27.00}{0.00}{10}
\emline{23.00}{3.00}{11}{23.00}{8.00}{12}
\end{picture}
\end{center}
Thus, the hexagonal lattice appears, which now may be imagined as made up of
its parts of the form%
\unitlength=0.50mm
\special{em:linewidth 0.6pt}
\linethickness{0.2pt}
\begin{picture}(8.00,8.00)
\emline{4.00}{0.00}{1}{4.00}{5.00}{2}
\emline{4.00}{5.00}{3}{0.00}{8.00}{4}
\emline{8.00}{8.00}{5}{4.00}{5.00}{6}
\end{picture}
. So, let us perform the
``star---triangle'' transformation
\begin{center}
\unitlength=0.5mm
\special{em:linewidth 0.6pt}
\linethickness{0.2pt}
\begin{picture}(27.00,8.00)
\emline{4.00}{0.00}{1}{4.00}{5.00}{2}
\emline{4.00}{5.00}{3}{0.00}{8.00}{4}
\emline{8.00}{8.00}{5}{4.00}{5.00}{6}
\put(10.00,4.00){\vector(1,0){7.00}}
\emline{19.00}{8.00}{7}{27.00}{8.00}{8}
\emline{27.00}{8.00}{9}{23.00}{0.00}{10}
\emline{23.00}{0.00}{11}{19.00}{8.00}{12}
\end{picture}
\end{center}
for each of them. One step of the evolution is over. It would be of interest
to reveal possible connections of this model with the model on the {\em
square} lattice from Section~\ref{dimers}.

Another interesting problem still unsolved: to describe the evolution
of the system in Section~\ref{gensys} in full, not up to gauge
equivalence.

\end{document}